**This is what a pandemic looks like: Visual framing of COVID-19 on search engines**

Mykola Makhortykh*, Aleksandra Urman** and Roberto Ulloa***

* Institute of Communication and Media Studies, University of Bern

** Social Computing Group, University of Zurich

*** GESIS – Leibniz-Institute for the Social Sciences

**Author Note**

This is the preprint version of the "This is what pandemic looks like: Visual framing of COVID-19 on search engines" chapter. The data collection for the chapter was sponsored from the SNF (100001CL_182630/1) and DFG (MA 2244/9-1) grant for the project "Reciprocal relations between populist radical-right attitudes and political information behaviour: A longitudinal study of attitude development in high-choice information environments" lead by Silke Adam (University of Bern) and Michaela Maier (University of Koblenz-Landau).

# This is what a pandemic looks like: Visual framing of COVID-19 on search engines

**Abstract**: In today's high-choice media environment, search engines play an integral role in informing individuals and societies about the latest events. The importance of search algorithms is even higher at the time of crisis, when users search for information to understand the causes and the consequences of the current situation and decide on their course of action. In our paper, we conduct a comparative audit of how different search engines prioritize visual information related to COVID-19 and what consequences it has for the representation of the pandemic. Using a virtual agent-based audit approach, we examine image search results for the term "coronavirus" in English, Russian and Chinese on five major search engines: Google, Yandex, Bing, Yahoo, and DuckDuckGo. Specifically, we focus on how image search results relate to generic news frames (e.g., the attribution of responsibility, human interest, and economics) used in relation to COVID-19 and how their visual composition varies between the search engines.

In today's "high-choice" (Van Aelst et al. 2017) media environment, search engines such as Google play an integral role in informing individuals about important societal developments. Together with social media and news aggregators, search engines increasingly become a major pathway to finding and engaging with the latest news (Newman et al. 2019) and historical information used to contextualize these recent developments (Zavadski and Toepfl 2019; Makhortykh, Urman and Ulloa 2021). Using complex algorithms, search engines filter and rank sources to counter the information overload and supply their users with the most relevant content. By prioritizing certain sources and judging how relevant a particular piece of information is, search engines serve as information gatekeepers (Laidlaw 2010) that structure individual and collective knowledge about the subjects varying from elections (Trielli and Diakopoulos 2022) to diseases (Paramita et al. 2021) to the matters of gender and race (Urman, Makhortykh and Ulloa 2022).

The gatekeeping functions of search engines translate into their substantial influence on different aspects of contemporary societies, in particular the ones related to information literacy (van Dijk 2010). This influence varies from search engines' information ranking affecting undecided voter preferences (Epstein and Robertson 2015) to their algorithmically constructed knowledge hierarchies reinforcing gender and racial stereotypes (Noble 2018; Urman and Makhortykh 2022). The functionality of search engines becomes particularly important at times of crises, when the need to create new hierarchies of knowledge in the face of uncertainty often leads to a state of "epistemic instability" (Harambam 2020) where the authoritative sources of truth are challenged, and alternative interpretations thrive. Under such conditions, filtering and ranking mechanisms used by search engines play an integral role in determining how individuals and societies understand the causes and the consequences of the current situation and decide on their course of action.

In this chapter, we look at how the above-mentioned complexities of information distribution via search engine algorithms affect pandemic literacy during the COVID-19 crisis. Understanding pandemic literacy as the ability to locate and effectively use information related to health threats associated with widespread epidemic diseases, we examine how search engines can facilitate, but also impede the ability of individuals and societies to stay informed about the COVID-19 pandemic. Specifically, we investigate the visual aspect of pandemic literacy by looking at visual representations of COVID-19 constructed by five major search engines in February 2020.

Our interest in COVID-19 is characterized by the profound epistemic crisis experienced in relation to the pandemic and amplified by the intense use of digital media for disseminating false information about the disease (Allem 2020). By scrutinizing how search engines represent the pandemic, we strive to advance the understanding of their role at the time of epistemic instability and how search engines construct knowledge hierarchies in relation to a new and highly contradictory phenomenon. Our decision to focus on the visual representation of COVID-19 by search engines as contrasted to its textual representation that was examined in earlier studies (e.g. Makhortykh, Urman, and Ulloa 2020; Toepfl, Kravets, Ryzhova and Beseler 2022) is attributed to images as an effective means of communicating complex phenomena, in particular those that are hard to express verbally (Bleiker 2018). Images also stir strong emotional responses that make them a potent catalyst of societal mobilization but also result in their frequent abuse for manipulating public opinion (Ruijgrok 2017).

To investigate how search engines represent and interpret COVID-19 visually, we combined qualitative analysis of visual frames—i.e., consistent patterns of selecting some aspects of the perceived reality and making them more salient (Entman 1993)—with quantitative techniques of algorithmic auditing used to investigate the functionality and impact of "decision-making algorithms" (Mittelstadt 2016). Utilizing a novel method of algorithmic auditing to extract visual search results, we compared the use of generic news frames (e.g., the attribution of responsibility, human interest, and economics; Semetko and Valkenburg 2000) for framing COVID-19 on five search engines: Google, Bing, Yahoo, Yandex and DuckDuckGo in English, Russian, and Mandarin Chinese. While doing so, we specifically investigated the following three research questions: How different/similar is the representation of COVID-19 via search engines compared with framing of earlier pandemics in legacy media? What types of news frames were prevalent in the representation and interpretation of COVID- 19? And are there substantial differences in the representation of COVID-19 among different search engines and languages?

**Theoretical background**

According to Gitlin (1980, 7), frames are "persistent patterns of cognition, interpretation, and presentation of selection, emphasis and exclusion" used to organize discourse about specific issues. Frames are often utilized as a conceptual device for studying how important societal developments are presented and interpreted by mass media and how they shape public opinion (de

Vreese, 2005). There are different typologies of frames, varying from the broad distinction between generic and issue-specific frames to more narrow differentiation among specific types of generic frames (for an overview, see de Vreese 2005). In our chapter, we utilize the typology introduced by Semetko and Valkenburg (2000) to distinguish five types of generic news frames (For a full description, see the methodology section.).

The process of framing involves selecting some aspect of the issue to make it more salient in order to promote its specific interpretation and/or treatment (Entman, 1993, 52). An example of such varying salience can be, for instance, two different groups of (visual) frames used to represent the same armed conflict: the first group of frames focuses on images of human suffering, thus stressing the human toll of warfare and the need for protecting civilians, whereas the other group emphasizes the images of smiling soldiers and military vehicles to present the conflict as "the good war" (Makhortykh and Sydorova 2017). The importance of framing explains its complicated relationship with the concept of information literacy: the effect of specific frames on individual perceptions of certain societal phenomena can be influenced by the presence/absence of specific information skills, as such skills can be essential both for the discovery of frames and/or their critical assessment.

A number of studies utilize the concept of framing to investigate how mass media represent health disasters, in particular the ones related to disease outbreaks (e.g., the recent H1N1 pandemic), thus stressing its importance for pandemic literacy. Pan and Meng (2016) use an issue-specific frame typology to study how the 2009 flu crisis in the US was framed and find varying degrees of prevalence for specific frames depending on the crisis stage (e.g., economic frames being more prevalent in the pre-crisis stage and medical issues being more visible during the crisis). Using a hybrid issue-specific/generic approach, Gadekar, Pradeep and Peng (2014) find distortions in the framing of H1N1 in India, in particular the unequal representation of actors' involvement, with a more supportive stance towards the government and more critical one towards the hospitals. Finally, Kee, Faridah, and Normah (2010), in their study of H1N1 framing in Malaysia, also deploy the typology of generic frames by Semetko and Valkenburg (2000) to identify the prevalence of the responsibility frame ( contrasting with the "least visibility" economic frame) as well as substantial consistency in the use of frames by different legacy outlets.

Current scholarship on pandemics framing, however, suffers two important gaps. First, the majority of studies so far omit visual representations of disease outbreaks and instead focus on textual content (e.g., news articles' text). Yet, visual content is an effective means of framing because of its more universal recognizability, compared with written texts, and strong potential for stirring emotional responses (Schwalbe and Dougherty 2015). Furthermore, the use of visual messages to communicate information is proven to have a substantial effect on individual behavior at the time of a pandemic (Updegraff et al. 2011), which is another reason for our interest in visual framing of COVID- 19.

Second, current research focuses on the ways pandemics are framed in legacy media, whereas its presentation via new platforms (e.g., search engines) remains understudied. Yet, digital platforms and their algorithms are increasingly recognized as influential actors in the process of news distribution that has far-reaching consequences for political and health-related information consumption (Entman and Usher 2019; Arendt 2020; Makhortykh, Urman and Wijermars 2022). Earlier research (Urman, Makhortykh, Ulloa and Kulshrestha 2022; Hannak et al. 2013) also demonstrates significant discrepancies in the output of search engines that can influence the quality of information received by their users, which is of particular concern in a time of emergency. By examining how search engines visually frame COVID-19, we strive to achieve a better understanding of the role of search engines in the process of framing, in particular their influence on the representation of health disasters.

**Methodology**

*Data collection*

To investigate how COVID-19 is framed visually via search engines, we collected data from five search engines: Bing, DuckDuckGo, Google, Yandex, and Yahoo. We selected these search engines because they are the ones with the largest share of the international search market (Statcounter 2020) or dominate their major local markets (e.g., Yandex in the case of Russia). To collect the data, we used a novel algorithmic auditing approach based on large-scale simulation of user browsing behavior via virtual agents (Ulloa, Makhortykh and Urman 2022). Virtual agents are software programs which can mimic user behavior (e.g., by entering URLs in the browser). Unlike earlier studies (e.g., Unkel and Haim 2019; Hannak et al. 2013) that rely on crowdsourced user data or small-scale simulations of browser behavior, our approach allows scaling the analysis to make it more robust and also permits conducting it in a controlled environment. The latter feature allowed us to investigate how image search results related to COVID-19 are filtered and ranked under a default - i.e., non-personalized - conditions.

The data were collected on February 26, 2020, two weeks before the World Health Organization declared COVID-19 a pandemic. We deployed 83 machines from the Amazon Web Services Frankfurt cluster with each machine hosting two virtual agents (one on Firefox and another on Chrome). The agents imitated three sessions of user browsing behavior and were evenly distributed among the search engines, so each search engine was queried by 33/34 agents. All agents started each session simultaneously and used image search for the term "coronavirus" in English (session 1), Russian (session 2) and Mandarin Chinese (session 3; simplified characters were used). Then, the agents navigated through the result page(s) to retrieve links to the first 50 images from HTML. Following each session, the browsers were cleaned to prevent previous searches from affecting the latter queries. We removed the data that can be accessed by the search engine's JavaScript (e.g., cookies) and the data accessible by the browser (e.g., browsing history).

*Data analysis*

To analyze collected data, we used qualitative framing analysis based on the generic frame typology developed by Semetko and Valkenburg (2000). The typology includes five types of frames:

1) conflict: frames emphasizing conflict between individuals, groups, or institutions; 2) human interest: frames bringing a human face or an emotional angle to the presentation of an issue/problem; 3) economic consequences: frames reporting an event or problem in terms of its economic consequences; 4) morality: frames putting the event or issue in the context of religious tenets or moral prescriptions; 5) responsibility: frames presenting an issue in such a way as to attribute responsibility for its cause/solution to either the government or to an individual/group.

Similar to Kee, Faridah and Normah (2010), we adopted a set of 20 questions used by Semetko and Valkenburg (2000) to measure the strength of the frames for each individual image we analyzed. Each question could be responded with either "yes" or "no"; the responses were used to measure the strength of the respective frame. The frame strength was calculated by summing the responses to the related questions (e.g., five questions for the human interest frame and the three questions for the economic one) which were quantified proportionally (e.g., each "yes" answer for one of the five human interest questions added 0.2 to the strength of the human interest frame for the respective image as it contained five questions, whereas for the economic frame questions each "yes" answer contributed 0.33 proportional to three questions).

For the analysis, we looked at the 50 most frequent images among the search results acquired by the agents querying the same search engine. These images were coded by the three authors of the chapter. The coding was then checked for consistency by one of the coders and the disagreements between the coders were discussed and consensus-coded.

**Findings**

The first finding is that **the number of actual frames in the search results is small compared with the usual number of frames used by legacy media** for framing the pandemics (Kee, Faridah, and Normah 2010; Pan and Meng 2016). As shown in Table 1, the ratio of frame strength is low for the majority of the search engines. In the case of search queries for Bing in English, we do not find any generic news frames at all, whereas in other cases, the number of images which can be treated as frames is extremely low. Instead of news frames in the usual sense of the word, many images prioritized by the search engines, in particular for English language, are schematic depictions of coronavirus particles. Examples of such images are shown in Figure 1 below. The major difference between such images is the color of the coronavirus particles and the level of the detail.

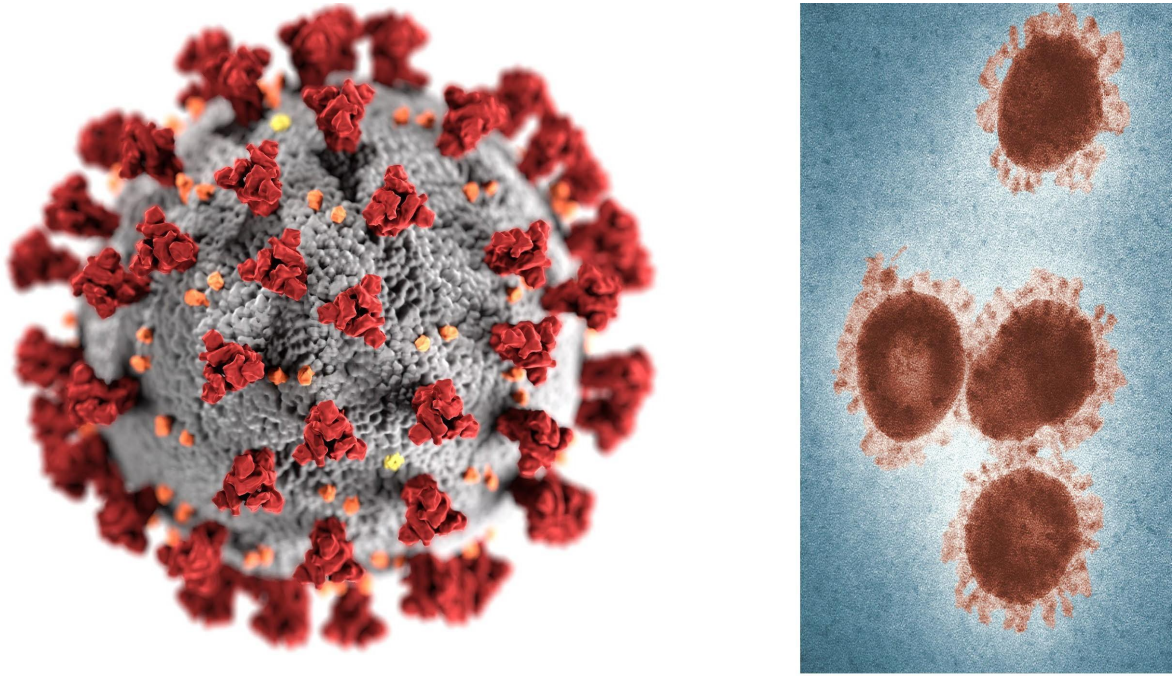

**Figure 1**. Schematic depictions of coronavirus search results. The images are taken from Unsplash and attributed to the Public Health Image Library from the Centers for Disease Control and Prevention.

The common feature of these non-frame images is their lack of interpretability that limits their use as a means of interpreting what coronavirus is. By themselves, the schematic depictions of coronavirus do not provide the user with information that is helpful for their understanding of the issue at hand. Not only are such images very different from generic news frames, but also their prevalence seems to fulfill only the most basic functionality of the search engine, i.e., finding the most specific images that illustrate the topic (e.g., virus particles in the case of COVID-19, which is not necessarily optimal for keeping users informed.

**Table 1**. Frame strength per engine-language combination (mean values)

|  | Responsibility | Human interest | Conflict | Morality | Economic |
|---|---|---|---|---|---|
| **Google-En** | 4.4 | 16 | 0.5 | 0 | 0 |
| **Google-Ru** | 2.8 | 20.8 | 0 | 0.66 | 0 |
| **Google-Ch** | 3.2 | 11.6 | 0 | 0 | 0 |
| **DuckDuckGo-En** | 0 | 0.4 | 0 | 0 | 0 |
| **DuckDuckGo-Ru** | 2.4 | 13.2 | 2.5 | 0 | 0.66 |
| **DuckDuckGo-Ch** | 0 | 2.4 | 0 | 0 | 0 |
| **Yandex-En** | 0.8 | 1.2 | 0 | 0.66 | 0 |
| **Yandex-Ru** | 2 | 1.6 | 0 | 0.66 | 0.66 |
| **Yandex-Ch** | 6.8 | 28.8 | 2 | 0.66 | 2 |
| **Bing-En** | 0 | 0 | 0 | 0 | 0 |
| **Bing-Ru** | 0 | 11.2 | 0 | 0 | 0 |
| **Bing-Ch** | 0 | 3.6 | 0 | 0 | 0 |
| **Yahoo-En** | 0.4 | 0.4 | 0 | 0 | 0 |
| **Yahoo-Ru** | 0.4 | 12.8 | 0 | 0 | 0 |
| **Yahoo-Ch** | 0.4 | 4.8 | 0 | 0 | 0 |

The second finding is **the substantially higher strength of the human-interest frame compared to the other four frames**. The most common appearance of this frame involves giving a human example or "human face" to the pandemic. Often, this representation of COVID-19 is supplemented by the emphasis on the pandemic's effect on individuals and societies. Usually, such frames picture masked medical personnel and/or the patients. In some cases (in particular, the queries in Chinese), it also includes schematic instructions explaining how COVID-19 is distributed in order to inform the public. An interesting case of the human-interest frame is observed for queries in Russian from Western search engines. There, the human-interest frame is dominated by visuals generating the feeling of empathy by showing images of felines, often with tears in their eyes or sitting on the hands of medics. The high visibility of felines can be related to search queries concerning the possibility of cats getting infected with COVID-19 which for some reason occurred particularly often in Russian.

Despite the high visibility of the responsibility frame in the case of legacy media coverage of other pandemics (Kee, Faridah, and Normah 2010), this frame occurs rarely in the case of COVID-19 framing on search engines. The other three types of frames are more underrepresented and appear just in a few engine-language combinations. This specific distribution of frames can be related to the differences in the framing process on search engines and legacy media, but also to differences in the framing of COVID-19 and earlier pandemics, so further research is required to clarify them.

**Table 2**. Cross-engine difference in frame strength (mean values)

|            | Responsibility | Human interest | Conflict | Morality | Economic |
|------------|----------------|----------------|----------|----------|----------|
| **Google**     | 3.47           | 16.13          | 0.17     | 0.22     | 0        |
| **DuckDuckGo** | 0.8            | 5.33           | 0.83     | 0        | 0.22     |
| **Yandex**     | 3.2            | 10.53          | 0.67     | 0.66     | 0.89     |
| **Bing**       | 0              | 4.93           | 0        | 0        | 0        |
| **Yahoo**      | 0.4            | 6              | 0        | 0        | 0        |

The third and the final finding points to **substantial differences in the framing of COVID-19 on various search engines and in various languages**. To examine these differences, we aggregated data on the average frame strength for each particular engine (Table 2) and each language in which we conducted searches (Table 3).

The cross-language comparison indicates that two groups of engines can be distinguished based on their framing of COVID-19. The first of them consists of Bing, Yahoo, and DuckDuckGo, where the number of frames in the search results is low and the results themselves are mostly represented by the schematic images of coronavirus particles (see Figure 1). The second group composed of Google and Yandex has more actual frames among the search results, in particular the ones related to responsibility and human interest. Furthermore, only engines from the second group include the morality frame. These differences can be attributed to several factors which are primarily related to the implementation of the ranking and filtering algorithms by different search engines. In the case of image search, these algorithms have different potential for recognizing the images and then contextualizing them to respond to the text queries. Some of these algorithms share substantial similarities (e.g., in the case of Bing, Yahoo and DuckDuckGo (Chris 2020)), whereas others (e.g., Google and Yandex) develop their unique solutions. Another explanation can be related to the differences in the type and the size of the audience utilizing the respective engines, which influences the quality of data used to filter and rank content for the respective queries.

**Table 3**. Cross-language difference in frame strength (mean values)

|  | Responsibility | Human interest | Conflict | Morality | Economic |
|---|---|---|---|---|---|
| **English** | 1.12 | 3.6 | 0.1 | 0.132 | 0 |
| **Russian** | 1.52 | 11.92 | 0.5 | 0.264 | 0.264 |
| **Chinese** | 2.08 | 10.24 | 0.4 | 0.132 | 0.4 |

Similarly, we observe a number of differences among the languages in which the queries were conducted. Specifically, almost all generic frames tend to be stronger in Russian and Chinese compared with English. The strength of the human-interest frame is particularly more pronounced with the queries in the former two languages returning more images of human actors in relation to COVID-19 as contrasted with more schematic depictions of the virus in English.

As in the case of cross-engine differences, the reasons for discrepancies among languages can be explained by several reasons. The stronger presence of the human-interest frame for Chinese queries can be attributed to China taking the strongest hit from the pandemic at the time of data collection (end of February 2020). Hence, for China, the human toll of COVID-19 was more visible compared with Western countries, where the number of infected was considerably lower. While in Russia, where the number of COVID-19 cases was also low, there seemed to be a disproportionate amount of interest in the consequences of the pandemic for house pets (in particular, cats), that together with the recognition of the emergency in areas neighboring China could contribute to the rise of awareness about the threat of COVID-19.

**Conclusions**

Traditionally, the process of framing is constituted by competition between actors, who compete "to define a problem, assign blame, and suggest who is responsible for addressing it" (Kee, Faridah, and Normah 2010, 107). However, in a high-choice media environment, where the audience frequently struggles with the information overload, this competition is increasingly affected by the algorithms (Entman and Usher 2018) that construct new hierarchies of knowledge and rank information sources. The importance of these new framing actors assumes particularly high importance under a state of epistemic instability, similar to that observed during the COVID-19 pandemic, where societies and individuals have to deal with the large volumes of (digital) information coming from different and often mutually contradictory sources.

Our findings indicate that visual images retrieved by the search engine algorithms do not necessarily have to do much with news frames in their traditional sense. While some of the search return results can be related to specific types of generic media frames (in particular, for Chinese and Russian language messages), in many cases these results are constituted by schematic images that do not necessarily provide a clear interpretation of the issue in question. These observations throw into question not only the framing potential of search engines but also their potential for increasing pandemic literacy, in particular in a time of emergency, where access to (visual) information can be important for individual and collective well-being.

In cases where search engines return news frames in relation to COVID-19, we find the human interest frame prevalent despite the more frequent use of the responsibility frame in earlier pandemics in legacy media (Kee, Faridah, and Normah 2010). The prevalence of the former frame can be attributed to the importance of bringing "a human face" (Semetko and Valkenburg 2000) to the representation of the disease and its consequences. The higher visibility of the human-interest frame can be viewed as a positive feature of framing COVID-19 as it can stress the severity of the emergency and importance of following healthcare prescriptions. At the same time, this specific frame can be argued to have the least interpretative value among the generic frames, instead appealing to audience emotions that might also enable the manipulation of public opinion.

Finally, we observed multiple cross-language and cross-engine discrepancies in the visual framing of COVID-19. These discrepancies do, to a certain degree, align with the "filter bubble" (Pariser 2011) argument by leading to rather different interpretations of the pandemic depending on the engine utilized by the users and the language of their search queries. Specifically, the probability for being exposed to the human interest frame in the context of COVID-19 (and, thus, being more aware of its severity and potential human toll) at the period when we conducted our study was higher for the queries conducted in Chinese, whereas the most common content for English queries was constituted by the schematic depictions of coronavirus.

It is important to note some limitations of the study. Because of our interest in the applicability of generic news frames to search engine-based framing, we used pre-established frame categories, not ones established via inductive coding as some earlier studies did. Our deductive coding could potentially limit our data interpretations, so in future studies we will go beyond the five generic frames we looked at currently. Another limitation is that the current study is based on a snapshot experiment conducted at the specific point of the COVID-19 pandemic. A more longitudinal approach is required to examine consistency of observed differences among the search engines and also changes in the framing of the pandemic through time. Additionally, it would be beneficial to rely on the broader selection of search queries and, possibly, use crowdsourcing for identifying what search combinations are the most used by the individuals searching for information about COVID similar to how it was done by Paramita et al. (2021). Yet, even with these limitations, our study raises important points about the role of search engine algorithms in the process of framing and their possible influence on the way the public is informed at the time of healthcare emergencies.